\documentclass[12pt]{article}

\def\b{\begin{equation}}
\def\e{\end{equation}}
\begin{document}
\title{The Day of Reckoning: The Value of the Integration Constant in the Vacuum
 Schwarzschild Solution} 
\author
{Abhas Mitra
\\
\normalsize{Nuclear Research Laboratory}\\
\normalsize{Bhabha Atomic Research Center, Mumbai- 400085, India}\\
Email: amitra@apsara.barc.ernet.in
}
\date{}

\maketitle



\begin{abstract}
The stongest theoretical support for Schwarzschild Black Holes (SBHs) is the existence of
vacuum Schwarzschild/Hilbert solution. The integration constant $\alpha_0$
in this solution is interpreted as the mass of the BH. But
by equating the 4-volumes (an {\em invariant}) associated with the Schwarzschild
metric and the Eddington-Finkelstein metric[1,2] for a SBH, 
 we directly obtain here the
stunning result that   SBHs have the unique
 mass, $M_0\equiv 0$! Thus the Event Horizon
(EH) of
a SBH ($R_g = 2M_0 =0$, $G=c=1$) gets merged with the central singularity
at $R=0$ and, after 90 years, the mysterious EH  indeed gets erased from the
non-singular  $R >0$ region of a completely empty ($R_0 =0$) spherical 
spacetime in accordance 
with the   intuition of the founders of General Relativity (GR)[3]. 
Consequently the  entropy  of SBHs have
the unique value of zero, which instantly removes the quantum mechanical ``information
paradox'' and the apparent conflict between GR and Quantum Mechanics, the two
pillars of modern physics[4]. And it is time to wonder how this simple result
was missed earlier and how the incorrect idea of (finite mass) SBHs took over
General Relativity almost 65 years back and then went on misleading Astrophysics,
Theoretical Particle Physics and even Quantum Information Theory! This clean result
firmly establishes the fact, as far as isolated bodies are concerned, General
Relativity is a singularity free theory even at the classical level! The Black Hole
Candidates with mass $M >0$ are thus not BHs and instead could be {\em hot} compact objects whose
possibility has so far been overlooked in favour of {\em cold} Neutron Stars or BHs.
This result is dedicated to the occurrence of 100 years of Relativity.

 \end{abstract}


Within two years of  formulation of General Relativity, Hilbert found the famous
spherically symmetric vacuum solution of Einstein equations for a ``Massenpunkt'',
i.e., a ``mass point''[3]:

\b
ds^2 = -\left(1 - {\alpha_0\over R}\right) dt^2 
+\left(1- {\alpha_0\over R}\right)^{-1} dR^2 + R^2(d\theta^2 + \sin^2\phi d\theta^2)
;~~ R \ge R_0=0
\e
where $\theta$ and $\phi$ are the polar angles and the {\em integration constant}
$\alpha_0$ is interpreted as twice the gravitational mass of the ``Massenpunkt'':
 $\alpha_0 
= 2M_0$ ($G=c=1$). Schwarzschild had found a similar looking solution where the radial
variable were $r$ instead of $R$: $R^3 = r^3 +\alpha_0^3$[3].
However, in a great disprivilege to Hilbert, this solution  got ascribed to 
Schwarzschild 
and
to avoid confusion  we too would continue to refer this solution as the
``Schwarzschild'' solution (SS). 
Because of Birchoff's theorem[2], this
vacuum solution also represents the exterior static spacetime of 
 a spherical body. If $M_0$ is the mass of this
spherical body, its value should depend on the value of $R_0$; i.e, $M= M(R_0)$,
$\alpha = \alpha(R_0)$ and, for $R_0 \le 2M$, let us again write:

\b
ds^2 = -\left(1 - {\alpha_0\over R}\right) dt^2 
+\left(1- {\alpha_0\over R}\right)^{-1} dR^2 + R^2(d\theta^2 + \sin^2\phi d\theta^2);
~~ R \ge R_0=\alpha_0=2M_0
\e

Also there could be a case when $R_0 > \alpha=2M$, and let in this case

\b
ds^2 = -\left(1 - {\alpha\over R}\right) dt^2 
+\left(1- {\alpha\over R}\right)^{-1} dR^2 + R^2(d\theta^2 + \sin^2\phi d\theta^2);
~~ R \ge R_0 >\alpha=2M
\e

For such a case, the apparent singularity
of the metric at $R=R_g=2M$ would be of no concern because the vacuum 
metric would cease to be
applicable in the region $R < R_0 $.

The original identification of $\alpha= 2M$ is done by matching the vacuum
solution in Metric (3) with the corresponding Newtonian solution at large $R$.

 Newtonian gravitation allows for the existence of a
spherical ``point mass'', i.e, $R_0 =0$, and since  in Newtonian gravitation, mass is essentially
of baryonic or leptonic origin (bare mass) with no negative ``dressing'' due to gravity or any self-energy, 
all masses including that of a ``point'' is necessarily finite, $M_0 > 0$.
And, in GR too,  it is has so far been {\bf assumed} that $M_0$ would 
continue to be finite
even when the spacetime is completely empty, i.e, $R_0 =0$ (mass point). 
It is this expectation which
gave rise to the  concept of Black Holes in the GR era. 
In the framework of this paradigm, 
 there would be real  physical vacuum spacetime not only for $R > 2M_0$ (metric 3)
but also for $R \le 2M_0$ (Metrics 1 and 2). 
And thus, in this paradigm, the singularity in metrics (1) and (2)

\b
g_{00} = \left(1 - {\alpha_0\over R}\right); \qquad g_{11} = \left( 1- {\alpha_0\over R}\right)^{-1}
\e
at $R =\alpha_0 = 2M_0$ is considered to be a mere 
coordinate singularity. Two reasons are often cited for the foregoing
assertion [1,2]:

(a)  GR requires that, in any truly non-singular region,  the determinant $g$
of the metric coefficients must be negative {\em in any coordinate system}
in order that its  sign  matches with the corresponding negative sign of
$g_{M}$ in a free falling Minkowskian spacetime:
\b
g = J^2 g_{M}
\e
where $J$ is the Jacobian of the relevant coordinate transformation. Since,
in spherical polar coordinates, $g_M = -R^4 \sin^2\theta$, we obtain
\b
g= - J^2 R^4 \sin^2\theta
\e  
Thus, at an arbitrary $\theta$, $g$ can vanish only at $R=0$ unless $J =0$ at some finite $R$,
a situation, highly unlikely, both, physically and mathematically.
Nonetheless, our conclusion 
{\em will not be based on} the
probable vanishing of $g$ at some region of spacetime.

For the diagonal metrics (1) and (2), one has
\b
g_1 = g_{00} g_{11} g_{\theta \theta} g_{\phi \phi}
 =-R^4 \sin^2 \theta =g_M
\e
Thus $J_1 =1$ despite the ``coordinate singularity'' and
 at $R=\alpha_0$, 
\b
g_1^{EH} =  -  \alpha_0^4 \sin^2 \theta = -16 M_0^4 \sin^2 \theta
\e
appears to be negative under the assumption $\alpha_0= 2 M_0 > 0$. 

(b) The Kretschmann scalar  associated with metrics (1) and (2) is
given by
\b
K = {12 \alpha_0^2\over R^6}
\e
and at the EH,
\b
K^{EH} = {12\over  \alpha_0^4} = {3\over 4 M_0^4}
\e
too appears to be   finite under the assumption
 $M_0 > 0$. 
 Now let us consider the Schwarzschild metric for a BH (i.e., metric 1 and 2)
in the isotropic coordinate
$\rho$ [1,2]:

\b
 R = \rho \left (1 + {\alpha_0\over 4 \rho}\right)^2;\qquad 
  \rho = {(R-\alpha_0/2) \pm \sqrt{R^2 -R \alpha_0}\over 2}
\e
in terms of which
\b
ds^2 = - \left({{1 - \alpha_0/ 4\rho}\over {1+ \alpha_0/ 4\rho}}\right)^2 dt^2  
+ (1 + \alpha_0/ 4\rho)^4 d\rho^2 + R(\rho)^2(d\theta^2 + \sin^2\phi d\theta^2)
\e
The corresponding metric determinant is
\b
g_2 = -  (1+ \alpha_0/4\rho)^2 (1- \alpha_0/4\rho)^2 R^4 \sin^2 \theta 
\e

and which does vanish at $\alpha_0 = \rho/4$, i.e, at
$R=\alpha_0$ (see Eq.[11]). This vanishing of $g_2$ at $R= \alpha_0 = 2M$ 
should immediately lead to an
introspection about the true nature of the ``coordinate singularity''. To this effect,
recall the obvious fact that the variable $ R$ has only two extrema; one at 
$R=  \infty$ and another at $R=0$ and correspondingly we must have
 $dR=0$ only at $R=\infty$ and $R=0$. 
But from Eq.(11), it follows that
\b
{dR\over d\rho} = (1- \alpha_0/ 4\rho) (1 + \alpha_0/ 4\rho) 
\e
and $dR=0$ not only at $\rho =\infty$, i.e, at$R=\infty$ and at $\rho = 
-\alpha_0/4$, i.e, at $R=0$, but also at $\rho=
+\alpha_0/4$, i.e, at $R=\alpha_0$. Hence $R$ has a
{\em  minimum  at $R=\alpha_0$ where $g_2=0$}!
Therefore,  $R=\alpha_0$
and $R=0$ must correspond to same spacetime where $R$ has its only minimum. And
this is possible only when $\alpha_0 =0$.

 Note that in case $R_0 > \alpha_0$,  Eqs.(11-14) will
cease to be valid for $R \le \alpha_0$ and we would not obtain the result
 $g_2 = dR=0$ at $R=\alpha_0$. These results are obtained {\em only when}
 we consider
$R_0 \le \alpha_0$. And for the BH, $R_0 =0$. 

Now consider the shape of the metric for $R=\alpha_0$, i.e, at  $\rho = \alpha_0/4$ (Eq.[11]):
\b
ds^2 \to +16 d\rho^2 + \alpha_0^2 (d\theta^2 + \sin^2\phi d\theta^2)\ge 0
\e
Thus, in no case, the metric is timelike at $R=\alpha_0$ which definitely
means that $R=\alpha_0$ is not a mere coordinate singularity but on the other
hand, a genuine spacetime singularity where the fundamental condition,
 $ds^2 <0$ for a
material particle (atleast at a non-singular region), would break down. To see the actual nature of $ds^2$ at $R=\alpha_0$,
we recall the BH metric in the so-called double null coordinate[2]:

\b
ds^2 = - \left( 1- {\alpha_0\over R}\right) du dv + R^2(d\theta^2 + \sin^2\phi d\theta^2)
\e
where

\b
u = t \mp R_* ; \qquad v= t\pm R_*; \qquad R_*=  R + \alpha_0 ~
\log \left({R\over \alpha_0} -1\right)
\e
The $\mp$ and $\pm$ signs in the above definition take care of both
ingoing and outgoing geodesics.
This shows that for a radial geodesic, as $R\rightarrow \alpha_0$,
 $ds^2 \rightarrow 0$ rather than $ds^2 >0$. Then from Eq.(15) it transpires
that as $R\rightarrow \alpha_0$, $d\rho \rightarrow 0$ and $\alpha \to \alpha_0
\rightarrow
0$, a result already obtained.  Physically, occurrence of $ds^2=0$ would mean that
for any infalling material particle, the 3-speed would be equal to the speed of
light, the maximum permissible speed in relativity.
Also note from Eq.(11) that {\bf only} when $\alpha_0 =0$, $\rho=R=$ {\bf real} over the entire manifold.
In the passing, note that the dererminant $g_3$ associated with the metric (16)
\b
g_3 = -  {1\over 4}\left(1 - {\alpha_0\over R}\right)^2 R^4\sin^2 \theta 
 \e
 too vanishes at $R=\alpha_0 = 2 M_0$. 

To ensure that  all the readers are fully convinced about the
resolution of this 90 year old mystery, we finally recall the Eddington-Finkelstein
metric[1,2] which was designed exclusively for SCHs, i.e, for 
$R_0 =0$, $M=M_0$, and $\alpha =
\alpha_0$:

\b
ds^2 = - \left(1 - {\alpha_0 \over R}\right)dt_*^2 \mp {2\alpha_0\over R} dt_* dR 
+ \left(1 +{\alpha_0\over R}\right)dR^2 + R^2(d\theta^2 + \sin^2\phi d\theta^2)
\e
where the Finkelstein time coordinate
\b
t_* = t \mp \alpha_0 \log \left( {R\over \alpha_0} -1\right)
\e
\b
R_* = R;\qquad \theta_* =\theta;\qquad \phi_* =\phi
\e

The corresponding metric coefficients are
\b
g_{{t_*} R} = -(1-\alpha_0/R), ~~ g_{RR}=(1+\alpha_0/R),~~ g_{{t_*} R}=g_{R t_*} 
=\alpha_0/R
\e 
In this case the determinant is same as $g_1$:

\b
g_4 = -g_{\theta \theta} g_{\phi \phi}(g^2_{t_* R} -g_{t_* t_*}
 g_{RR})
= -R^4 \sin^2 \theta = g_1=g_M
\e
Again, everywhere, $J_4 =1$. Now let us apply the principle of {\em invariance of 4-volume} for the coordinate
systems ($t$, $R$, $\theta$, $\phi$) and ($t_*$, $R$, $\theta$, $\phi$) 
 at arbitrary $R$ and {\bf not necessarily at $R \le 2M$}:
\b
\sqrt{-g_4}~ dt_*~ dR~ d\theta ~d\phi = \sqrt{-g_1}~  dt~ dR~ d\theta~ d\phi
\e
Since $g_4= g_1$, we promptly obtain
\b
dt_* = dt
\e

By using Eq.(20) in the foregoing Eq., we find

\b
dt \mp {\alpha_0 \over R - \alpha_0} dR = dt
\e
which leads to
\b
  {\alpha_0 \over R - \alpha_0}  = 0; \qquad at~any~ R
\e

Thus, {\em in a most direct manner}, we obtain the stunning result that the mass of
the Schwarzschild BHs:
\b
M_0 = M(R_0=0) = \alpha_0/2 \equiv 0
\e
Prosaically, this means that,  the gravitational mass of a ``mass point''
($R_0 =0$)
is {\bf actually} $M_0\equiv 0$ which, however, following the Newtonian hangover has so far been assumed
to be finite!

 Then  the EH merges with the central singularity at $R=0$ and
hence the metric (1) has only one singularity, the 
singularity  at $R=\alpha= \alpha_0=0$.  Note that even when
$\alpha = 2M=0$, $K^{EH} = K (R=0) =\infty$ implying that this is a
curvature singularity. It may be also recalled that in GR, unlike in
Newtonian gravitation, occurrence of $M=0$ does not necessarily mean
absence of matter. A zero total energy occurs whenever all sources of
positive energy like baryonic or leptonic mass energy and internal energy
get exactly balanced by negative self-gravitational or any other self
energy.
Now let us briefly recapitulate the series of wrong notions which led to
the greatlt incorrect idea of finite mass SBHs:

1. The integration constant $\alpha_0$ appearing in the 
Hilbert  (now known as Schwarzschild) solution was erroneously considered to be
positive definite ($\alpha_0 >0$) when attempt should have been made to fix it
with suitable physical condition(s), as is the norm for fixing integration constants.

2. Even assuming that $\alpha_0>0$, one could have attempted to find out the 3-speed
($v$) of a test particle at $ R=\alpha_0$ and $ R <\alpha_0$ and by realizing that once
$v \to 1$ in any coordinate system, rule of relativistic addition of velocities must ensure
that $v\to 1$ in all other coordinate system.

3. The fact that the acceleration and more importantly, physically measureable
{\em Acceletation SCALAR}, $a$, blows up at $R=\alpha_0$[3, 5] was always ignored. Attention
was focussed only on the geometrical scalars such as Kretschmann scalar without 
realizing
that at a mere coordinate singularity none of the
 Physical Scalars would vanish. On the other hand, the value of $\alpha_0$,
an integration constant, 
should have been fixed by taking help of physical quantities, such as $a$.

4. Chandrasekhar's discovery of an upper mass limit for White Dwarfs only gave
the upper mass limit of {\bf cold} static objects and all it meant is 
that collapse can proceed beyond
the cold White Dwarf stage. But this
 was interpreted as existence of SBHs as the immediate next step.

5. Similarly in context of the existence of the Oppenheimer-Volkoff limit[6], which at present, is
considered to be $\sim 3-4 M_\odot$, it was ignored that this limit corresponds to
{\bf cold} and strictly static baryonic objects. The possibility of likely existence
of Einstein collapse solutions for {\bf hot} (i.e, objects primarily supported by
trapped radiation pressure) and quasistatic objects (which would generate pressure
gradient even without any nuclear fuel) was never considered.

6. We recall that there is only one {\em exact analytical} solution of spherical
gravitational collapse where an uniform dust of mass $M$, initially ($t=0$) 
at rest with a radius $R_i$, collapses to a SBH in a proper time[7]
\b
\tau_c = {\pi\over 2} \left({R_i^3\over 2 GM}\right)^{1/2}
\e
Since the dust ball is at rest at $t=0$, we can use the equation for hyrdo-static
balance[6] at $t=0$:
\b
{dp\over dR} = - {p +\rho_i\over R(R-2M)} (4\pi R^3 p + M)
\e
Since for a dust $p\equiv 0$,
 we have $dp/dR \equiv 0$, and then the foregoing Eq. yields
$\rho_i =0$. From thermodynamical point of view too, whether at rest or not,
a $p=0$ equation of state is physically obtained only if $\rho =0$. Therefore,
trivially, the mass of the dust ball is zero for a finite $R_i$. Hence the mass of the
resultant SBH is indeed $M=0$. 
But this was always ignored, it was pretended that eventhough, $p=0$, we must have
$\rho =finite$ and $M$ too would be finite in tune with the idea that the integration
constant $\alpha_0 >0$ (when it was actually zero).

And when $M=0$, from Eq.(29), it follows that,
$\tau_c =\infty$. Thus though the $M=0$ SBHs are, mathematically, allowed
by GR, they  cannot be realized in an universe with finite proper age.
 
Since there is no EH in a finite proper time, there is no trapping or loss of
quantum information either and hence
there is no question of any conflict between GR and quantum mechanics[4]. 

In some  different papers, it was shown that even for the most general case of spherical
collapse (i.e, not necessarily for uniform dust), no trapped surface is ever formed[8,9,10].
Thus a collapsing fluid must always radiate and if it would be assumed to undergo
continued collapse $M\to 0$ asymptotically. This is the reason  that the integration
constant $\alpha_0 = 2M_0$ turned out be identically zero. Hence the observed BH candidates with masses often much higher
than the upper mass limit of {\em cold baryonic} bodies in hydrostatic equilibrium,
cannot be SCBs. Detailed analysis of recent observational data indeed suggests that
the BH candidates have strong intrinsic magnetic moments rather than any EH[11,12,13,14].
 And it
has been suggested that the BH candidates are Magnetized Eternally Collapsing Objects
(MECOs). These are extremely {\em hot} objects in
quasistatic equilibrium due to
extremely strong radiation and magnetic pressure[10,11,12,13]. However, if Quantum Gravity
would be invoked, the supposed Black Holes could be {\em cold} and {\em static} configurations
with hard surfaces[14].

Much investigations would   be required to be certain about the precise nature of the observed BH candidates.
 In any case, since the BH paradigm was built on the {\em assumption}  $\alpha_0 >0$,
and now that it is found that, actually, $\alpha_0 =0$, this paradigm collapses immediately irrespective
of the fact that, at this moment, most of the authors would close their eyes to ignore
this stunning result. With $\alpha_0 =0$, the Schwarzschild, Hilbert and related solutions
become identical and the physical confusion arising from mathematical gauge freedom vanishes for the vacuum case.




\end{document}